# The medium is the message:
# How neurometabolic entanglement generates conscious experience

Terrence W. Deacon[1], Miguel Garcia-Valdecasas[2], Michael Jacob[3]

**Short Abstract:** We describe a process of neuro-metabolic entanglement (NME) that demonstrates why conscious experience cannot be explained in terms of neural information processes alone but must include neurometabolic substrate factors. Central to this account is a recursive, Möbius-like loop of neural connections that entangles patterns of forebrain neural information processing with midbrain regulatory systems that modulate forebrain metabolism. This enables forebrain neural activity to indirectly modify the global distribution of forebrain metabolic resources. It also constitutes a bottleneck that integrates diverse neural processes into a unitary experience and explains why so many psychoactive chemicals affect midbrain regulatory neurotransmitter systems.

**Long Abstract:** Most neuroscientific accounts of consciousness assume that conscious experience is constituted by the information conveyed between neurons. The physical-chemical details of the metabolic and physiological processes that convey this information are correspondingly presumed to be relevant only insofar as they support the organization of information processes. We describe an alternative perspective that denies this presumed independence of the processing of neural information and the metabolic-physiological details of these processes. We call this alternative paradigm the theory of neuro-metabolic entanglement (NME). NME proposes that mental experience exemplifies the inseparability of the organization of neural information processes from the constraints imposed by the material-energetic properties of the neural medium itself. Central to this account is a recursive, Möbius-like loop of neural connections that entangles patterns of forebrain information processing with midbrain regulatory systems that modulate forebrain metabolism. This enables patterns of neural activity to indirectly modify the global distribution of metabolic resources underlying these information processes. It also constitutes a bottleneck that integrates diverse neural processes into an adaptively relevant unitary experience organized in terms of the global redistribution of metabolic support. Computational models of neural function cannot treat energy supply as a fixed background parameter; our proposed Möbius-like loop of neuro-metabolic entanglement offers a framework for future models in which neural activity is dynamically and recursively modulated by the network's own metabolism. Recognizing this entanglement is a first step toward dissolving the explanatory gap highlighted by the so-called "hard problem" of consciousness.

---

[1] University of California, Berkeley, Department of Anthropology
[2] Universidad de Navarra, Spain, Department of Philosophy
[3] University of California, San Francisco, Department of Psychiatry and Behavioral Sciences

## 1. Introduction

The majority of efforts to explain how brain processes give rise to conscious experience attribute it to the patterning of the signals propagated from neuron to neuron. Although most neuroscientists would argue that the felt quality and content of conscious experiences are functions of neural signal processing, exactly how and why this neural activity gives rise to experience remains mysterious. The persistence of this explanatory gap has led some philosophers to question whether consciousness can be fully explained in physical terms and has been called the "hard problem of consciousness" by philosopher David Chalmers (1996). Even though cognitive neuroscientists and philosophers may differ on the implications, this formulation of the problem reflects a general perspective that is broadly accepted (Francken et al. 2022).

The assumptions that underlie this perspective can be superficially described as follows: Conscious experience is constituted by neuron-to-neuron information processes where the physicochemical details of the processes conveying this information are relevant only insofar as they support these information processes. In simple terms, this assumes that the physiological details of the neuronal medium are no more determinative of the form of conscious experience as are wires and electric currents of telephone circuits determinative of what is communicated through them.

In this paper, we argue that it is time to abandon this information-centered framing and with it, the narrow range of neurocognitive theories that it allows. In its stead, we introduce a theory that we call *Neuro-Metabolic Entanglement* (NME). It proposes that conscious experience —the subjective *feel* of neural processes— is constituted by the way that the information processes of the brain are influenced by physiological changes of its neuronal medium and how this physiological medium is in turn recursively influenced by the information processes that it supports.

Thus, instead of treating the physiology of brain processes like a channel that remains unmodified by the information it conveys, and treating the information as unmodified by the physiological properties of the neurological medium, we argue that conscious experience can only be understood in terms of their inseparability and causal entanglement.

In what follows, we trace the anatomical and physiological details of this entanglement, demonstrating how the neuro-metabolic contributors to brain function (e.g. blood flow, neuronal excitability, plasticity, etc.) modify patterns of neural information processes, and vice versa. Paraphrasing an enigmatic aphorism by Marshall McLuhan, NME theory posits that "the medium is the message" to the extent that mental experience emerges from the entanglement of neural information processes with their own physiological embodiment.

## 2. Unconscious neural activity

Neurons never sleep. Like Alice in *Through the Looking-Glass,* neurons need to be constantly active just to stay the way they are (Raichle 2015). This is because neurons must constantly pump ions across their cell membranes to maintain an ionic imbalance. This ionic imbalance can be disrupted by signals from other neurons. When synaptic inputs push the membrane potential beyond a critical threshold, this ionic potential is suddenly discharged, generating an action potential that propagates out along its axon to stimulate other neurons. But being close to this threshold also causes neurons to fire spontaneously with some characteristic regularity (Uddin 2020). Because each neuron is interconnected with hundreds or even thousands of others, local networks tend to exhibit ongoing patterns of collective oscillatory activity. This incessant activity helps explain why, even at rest, the brain requires roughly ten times more metabolic support than most other organs (Raichle and Gusnard 2002; Magistretti and Allaman 2015; Müller et al. 2013, Jacob et al. 2023).

This spontaneous local neural network activity is critical for managing body functions. In each second, the brain is flooded with inputs from the senses: from touch, sound, vision, taste, and odor receptors, as well as from signals relaying bodily needs such as hunger and respiration. The brain must continuously track these changing patterns of stimulation originating both within and outside the organism and adapt to match this input.

Yet nearly all of this vast and incessant activity seldom rises to the level of conscious awareness; it only becomes available to conscious attention when an unexpected change occurs. This is not because the inputs fail to have an effect or are somehow inhibited. Indeed, they are constantly anticipated by spontaneous neural activity.

Consider the familiar experience of "tuning out" background noise, for example, a phenomenon often described as habituation. A steady background hum in an air-conditioned room quickly fades from awareness, yet when the air conditioner shuts off, the resulting silence immediately captures one's attention. This is because the auditory system adapts to redundant sensory stimuli by generating a matching pattern of brain activity. This gradual fading from conscious awareness enables the brain to minimize metabolic expenditure. While expected or redundant inputs require relatively little processing, unexpected events demand additional neural activity. Thus, the degree of conscious awareness appears to be inversely correlated with the level of predictive redundancy: stimuli that conform closely to expectations tend to recede from awareness, whereas violations of expectation become salient.

The ubiquity of this spontaneous, unabated anticipatory brain activity, is evidence that neuronal information processing alone cannot be the basis of conscious experience. For the most part, neural information processing takes place unconsciously, and unconscious neural activity is the default state. As a result, consciousness is the exception. One might even describe the function of consciousness as a means to find ways to do things unconsciously (Poldrack et al. 2005). It appears to get turned on in order to turn itself off.

This is one of the most pervasive features of brain function: its *anticipatory* organization. In many respects, this spontaneous predictive function of neural circuits provides one of the most basic organizing principles of brain function. Theories of allostasis (Theriault et al. 2025; Seth and Bayne 2022) track the predictive regulation of the body's energetic and physiological needs and constitute much of this anticipatory work. Its centrality has been recognized by several theories of consciousness. One of the most prominent is the Free Energy Principle (FEP) associated with Carl Friston (Friston 2010). Drawing on an analogy with the thermodynamic tendency of far-from-equilibrium systems to minimize free energy (as in Gibbs free energy in statistical mechanics), Friston proposes that deviations from anticipated neural inputs (which he describes as "surprisal") track levels of informational and biological changes that require the organism's attention. But a key question remains: why should deviation from expectation be subjectively felt?

We think that this deviation can be better explained by taking the thermodynamic analogy more literally.

Consider the well-known correlation between mental experience and the measures of metabolism that are the basis of fMRI (functional Magnetic Resonance Imaging). The interpretation of fMRI data is based on evidence showing that a regional increase in blood flow and the use of oxygen and glucose is correlated with an increase in baseline neural metabolism in that region. The resulting increase in oxygen and glucose delivery presumably compensates for this increased neural activity. Without this metabolic response, neural responsivity to inputs will be constrained and irregular. This interaction between neural input, metabolic support via blood flow, and neural output firing patterns is illustrated in **Figure 1.**

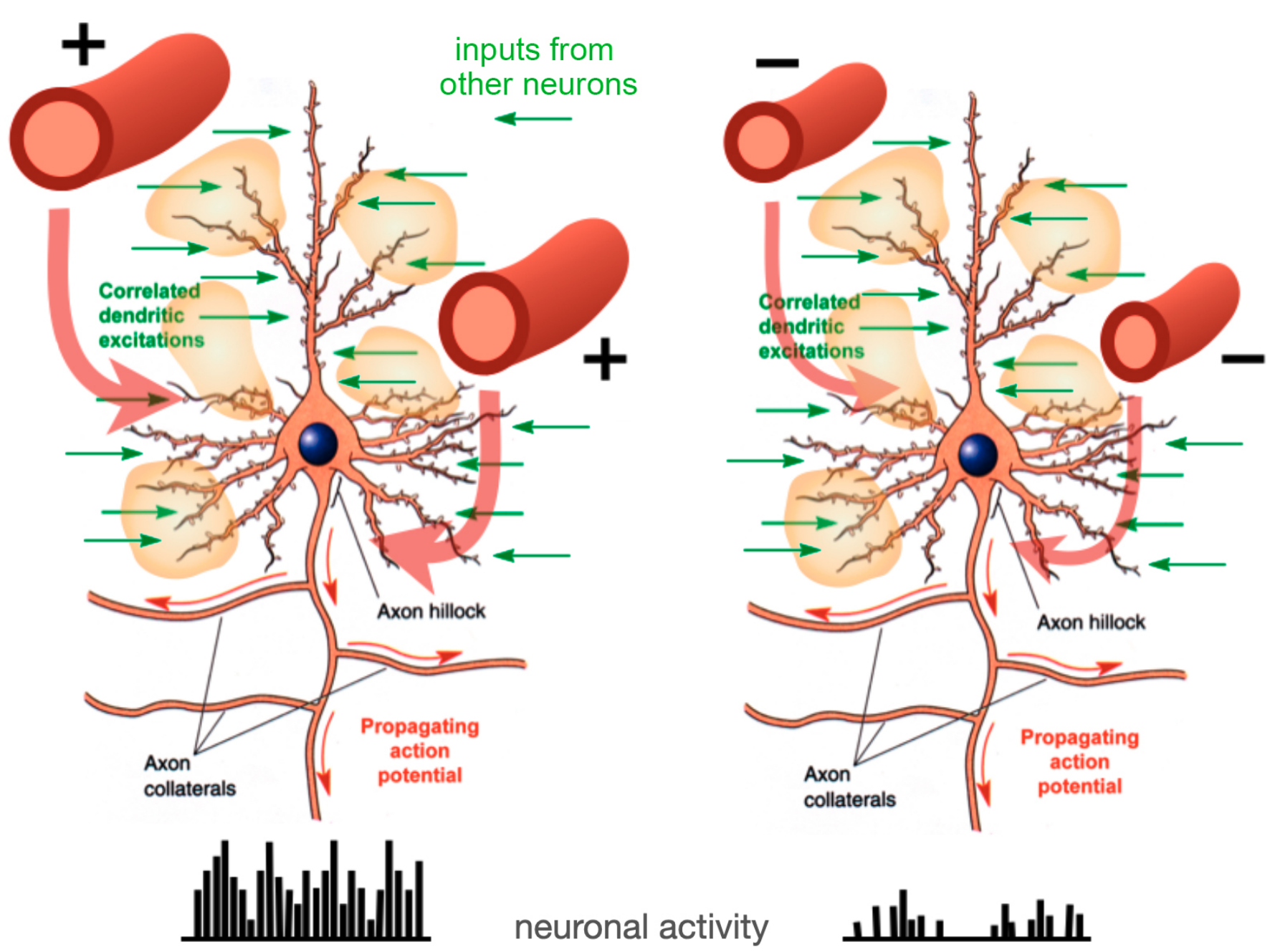

**Figure 1.** A depiction of pyramidal neurons in response to different metabolic conditions but with similar neural inputs. The left image depicts neural input-output relations with unrestricted energy supply and blood flow. The black bar graph below the axon output indicates a high regular firing rate. The right image depicts neural input-output relations with restricted energy supply and reduced blood flow due to capillary restriction. The black bar graph below its axon indicates a significantly reduced and irregular firing rate. Increased neural activity due to increased neural signal input can locally induce an increase in local blood flow, but metabolic support can also be modulated independent of local activity with the same functional consequences. Thus, changes in metabolic support can affect the pattern of the information generated by neural activity.

The balance between neural activity and its physiological support is continuously perturbed by fluctuations in neural activity and continually reestablished through metabolic and vascular regulation. However, this coupling is subject to significant temporal mismatch that reflects a fundamental disparity in timescales. Neural activity propagates through densely interconnected networks within milliseconds, whereas the hemodynamic responses that adjust local blood flow and replenish metabolic resources develop over several seconds (Hillman 2014; Drew 2019). Hence, metabolic support does not instantaneously track neural demand; instead, there is a continually shifting tension between neural signaling and the slower processes that sustain it (Lecrux et al. 2019; Meyer-Baese et al. 2025; Howarth et al. 2021).

Unlike FEP predictions, the subjective *feel* is not just an informational "surprisal" but the entangled metabolic work this necessarily involves. Since the introduction of fMRI, it has been known that fluctuations in the BOLD signal closely reflect patterns of neural activity (Logothetis and Wandell 2004). Focused perceptual paradigms demonstrate that conscious, subjective report tracks the magnitude of the BOLD response those sensory regions (Hesselmann and Malach 2011; Wohlschläger et al. 2016), even as extended activation across multiple regions is implicated in conscious awareness from meta-analytic studies (Bisenius et al. 2015). Critically, BOLD responses are generally measured relative to a baseline (Buxton 2012), emphasizing that this signal is always active (see Figure 2).

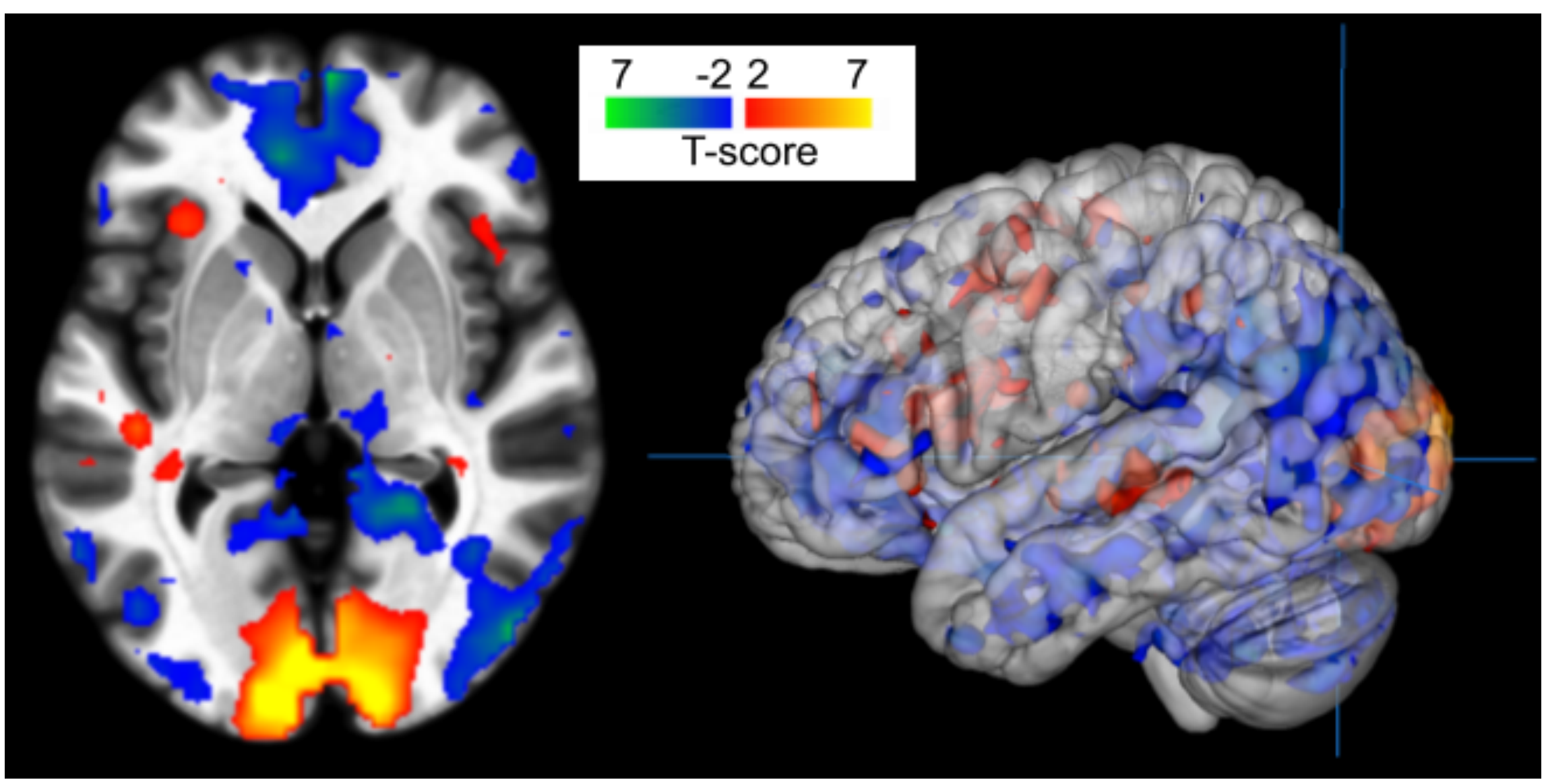

**Figure 2.** Statistical parametric maps of task-related blood-oxygen-level-dependent (BOLD) signal change during a semantic priming task. Activity reflects the change in BOLD signal relative to baseline to a non-matching, semantic stimulus, averaged across 27 healthy individuals. Activity is shown on an axial slice of a template brain (left, z=0) and a semi-transparent three-dimensional glass-brain rendering in lateral view (right). Voxel color reflects the T-score (magnitude of deviation from baseline) encoded by the scale bar: warm colors (red–orange–yellow, T = +2 to +7) denote significant relative activation (increased BOLD signal) and cool colors (blue–green, T = −2 to −7) denote significant relative deactivation (decreased BOLD signal) in the non-match condition relative to baseline. Prominent positive activation is seen in the bilateral occipital (visual) cortex, with widespread negative T-values across frontal, temporal, parietal, and subcortical regions. Crosshairs on the rendering indicate the reference coordinate. Data from (Jacob et al. 2019).

Most fMRI researchers treat the BOLD response as a surrogate for neural-electrical information processing in the same brain region, and by assumption, the heightened involvement of that brain region with some parameter of the task that the fMRI subject is engaged in.

## 3. Why mental effort feels like work

NME theory begins with the assumption that patterns of local neural network activity are self-reinforcing and resist perturbation in order to reliably anticipate redundant input. NME theory adds that one means to overcome this functional inertia is to extrinsically modify the metabolism that supports it. This needs to be accomplished indirectly. Redistributing metabolic resources also requires physiological work. Modifying the delivery of metabolites or changing cellular physiology are physical processes that involve modification of material substrates. We argue that the imposition of this physiological work is what is experientially *felt* as mental *effort,* not just the change in the pattern of neural activity that results.

One prediction of this inversion of perspective is that the intensity of conscious experience should vary in correlation with the level of metabolic work involved. This might be tested by methods able to measure experiential intensity as it correlates (or not) with the deviation from baseline metabolic and hemodynamic signals. Similarly, states such as heightened arousal, shifts in alertness, and states of mental fatigue should also be related to the level of physical work involved.

In general, this suggests that conscious awareness arises whenever the brain's anticipatory inertia is overcome, or when metabolic resources are required to deviate from some prior persistent activity. The greater the resistance to such deviation —the more work required to shift between anticipatory states— the more vivid and foregrounded the resulting conscious experience. This helps to explain why novel, unexpected, or startling events are experienced more vividly than familiar, predicted, or habitual ones. It is not because novelty is intrinsically rewarding or threatening, but because it requires a greater metabolic investment to deal with the functional reconfiguration of brain patterns that novelty requires. It also explains why what it *feels like* to be conscious is not an on-and-off property but a *graded* feeling. This is what should be expected if it reflects the amount of physical work required to modify some existing activity. The greater the resistance to change, the more work is needed. And the more work, the stronger the corresponding feeling of mental effort.

Taken together these considerations lead us to see conscious awareness as being in a *negative* relationship to the brain's normal, continuous, largely unconscious anticipatory activity. It is the special case.

This still leaves two important aspects of the problem of consciousness unaddressed. First, identifying the intensity of the felt effort involved in difficult mental activity with the underlying physical work does not explain how the informational and metabolic aspects of the neural medium become entangled. Second, even if the basis of this subjective quality is specified, the more fundamental question of the locus of the self—that is, the experiencer—remains unresolved. We address these two issues in the next two sections, respectively.

## 4. A Möbius twist

One of the most widely cited theories proposing how brain organization contributes to phenomenal consciousness is Giulio Tononi's Integrated Information Theory (IIT) (Tononi et al. 2016) (Tononi 2004; Tononi 2008; Oizumi, Albantakis & Tononi 2014). IIT postulates that the degree of consciousness experienced by an information-processing system such as the brain depends on the extent to which its components form an irreducibly integrated causal network. Specifically, IIT proposes that a system in which each component's state both influences and is influenced by the states of other components will exhibit a degree of consciousness proportional to the system's level of integration.

This line of reasoning implicitly assumes that consciousness is a function of causal recursion: the more intensive the causal recursion, the more that every part of a system depends on every other part. There is an intuitive attraction to this association. The very notion of consciousness has a recursive self-referential sense. And yet, although recursive processes contribute a degree of system integration and closure, they do not in themselves define a self/non-self distinction.

So, although some high level of functional and causal integration of neural signaling is widely regarded as a *necessary* condition for consciousness, it is unclear why this should be taken to constitute its *sufficient* condition. What appears to be missing is a form of recursion that refers to its own concrete individuation, not merely certain of its formal attributes.

To address these questions, consider an analogy: the curious geometry of a Möbius loop (shown in the middle of Figure 3). This simple strip of paper has fascinated children and adults alike because it appears to erase the distinction between opposites. Although the two sides are necessarily distinct and mutually exclusive at every point, following either surface of the loop reveals that they are continuous with one another.

Now consider the multi-step circuit of neuronal connections depicted in Figure 3. In three synaptic steps, a loop of neural connections can be traced from cerebral cortex to structures in the basal forebrain to nuclei in the midbrain and brainstem, then back to cortex. Neurons in the cerebral cortex receive and integrate information from the external world and initiate motor activity. Besides many other targets, some cortical neurons extend their axon to synapse on neurons in structures around the basal midline of the forebrain. The structures include subcortical nuclear structures—including the nucleus accumbens, the amygdala, and the hypothalamus. Among other functions, neurons in these structures assign positive or negative valence to these inputs. These forebrain structures then extend axons to contact neurons in specialized nuclei in the midbrain and brainstem. These include the locus coeruleus, the raphe nuclei, and the substantia nigra. These deep brain nuclei, in turn, extend axons back to the cerebral cortex, terminating across nearly all cortical areas, as well as many subcortical systems of the forebrain (Morgane et al. 2005; Pessoa 2024).

The neurotransmitters expressed by neurons in each of these nuclei are commonly referred to as regulatory neurotransmitters, and many of their projections to the forebrain do not even terminate on synapses but are released by axonal varicosities into intercellular space. In addition, many blood-borne metabolites (hormones, nutrients, carbon dioxide levels, and other molecular signals) that correlate with body physiology modulate the activity of neurons in these nuclei (Özçete et al. 2024).

Locus coeruleus neurons, in particular, are relevant to cortical metabolism because their axonal projections extend throughout the cortex and play a role in the regulation of local cortical blood flow. Norepinephrine, the neurotransmitter molecule released by locus coeruleus axons, influences the regional redistribution of cortical blood flow by activation of cells called pericytes that wrap around capillaries and contract to reduce blood flow. Similarly, widely projecting axons from neurons in the raphe nucleus and substantia nigra —whose primary neurotransmitter molecules are serotonin and dopamine, respectively— also modulate metabolic properties of target neurons, rather than conveying specific signal patterns (Korte et al. 2023; Sun et al. 2026; Giorgi et al. 2020; Mu et al. 2026). In this way, these midbrain-brainstem systems affect forebrain information processes indirectly, by up- and down-regulating the metabolic parameters of the neuronal medium, rather than directly contributing new neural signal information.

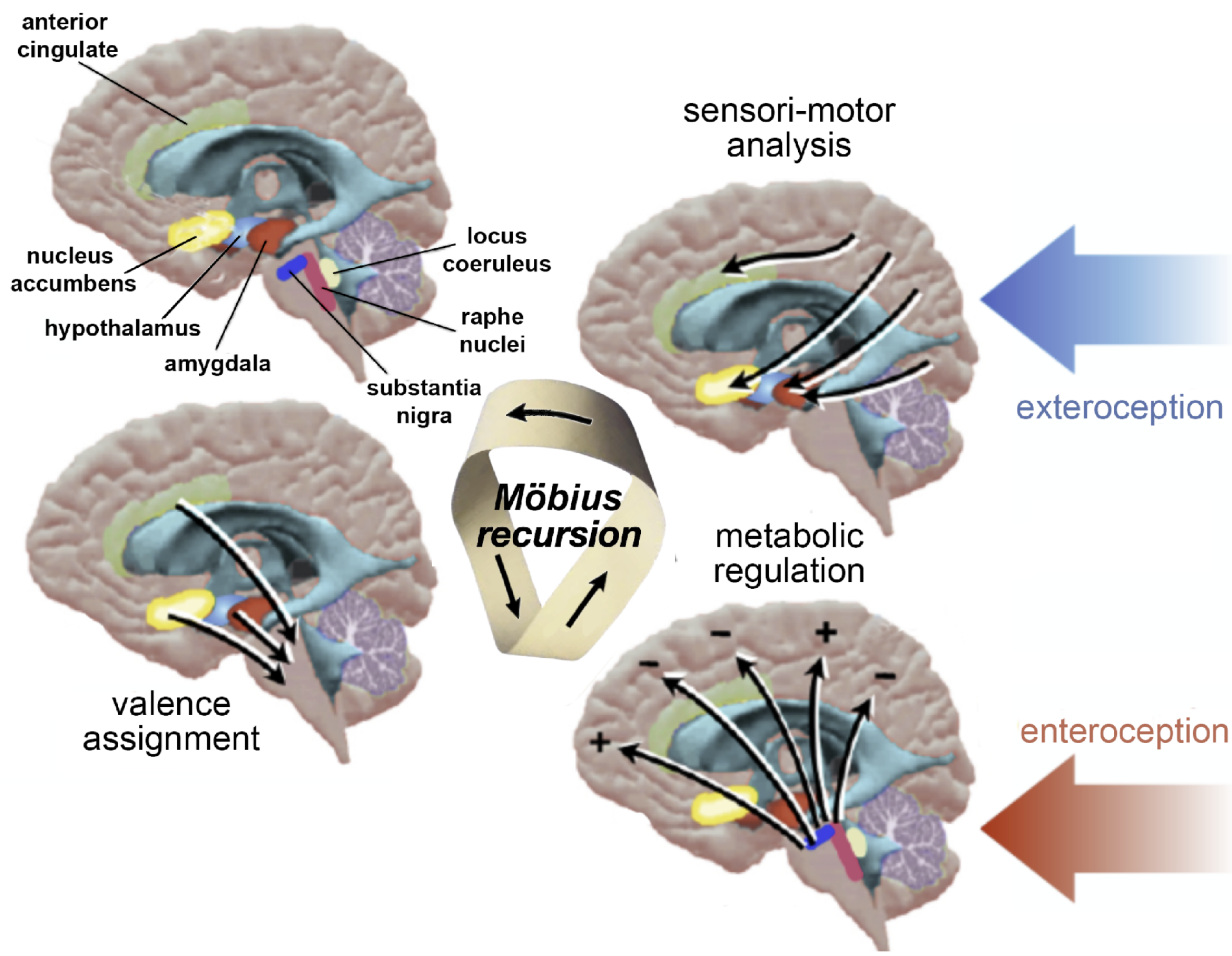

**Figure 3.** The circle of neuronal connections by which patterns of global cerebral cortical activity indirectly modulates the redistribution of local metabolism in the cerebral cortex. The top left image identifies some of the major subcortical nuclei involved in this loop of connections. This recursive loop of connections is diagrammed using the three brain images on the right. The three steps are: 1. output from cortical signals "assessed" by basal forebrain structures (e.g. nucleus accumbens, amygdala, and hypothalamus); 2. outputs from these structures are relayed to midbrain-brainstem nuclei (e.g. raphe nuclei, substantial nigra, and locus coeruleus); and 3. outputs from these midbrain-brainstem nuclei project throughout the forebrain to regulate multiple aspects of neural metabolism (e.g. distribution of blood flow, modification of neuronal response thresholds, changes in synaptic plasticity, etc.).

This recursive loop of neural connections can roughly be analogized to a Möbius loop because of the way it exemplifies how opposite but complementary attributes can nevertheless also be continuous with one another. The functional *twist* in this brain circuit occurs as the midbrain-brainstem nuclei project back to the forebrain structures. These projections do not primarily relay sensory, motor, or any form of structural information to synapses of forebrain neurons. Instead, the outputs of neurons in these nuclei mostly exert broad modulatory influences that generically modify the excitability, synaptic plasticity, and local energy available to these target neurons. By analogy, the neural message "side" of the loop becomes the physiological

"medium" side of the loop and vice versa as signals propagate around again. This continuous process recursively entangles neural message processing with neuronal medium properties.

Through this recursive relationship, neuronal informational processes can both be distinct from and at the same time continuous with the neurophysiological medium they are expressed within. Thus, what appears from one perspective as two distinct brain processes become phases of a single, unified process. Metabolism and neural signaling are not just correlated, but so recursively entangled that their respective functional effects become inseparable.

## 5. The metabolic bottleneck

The NME has implications for what is often described as the "binding problem:" i.e., the unitary nature of perceptual experience and conscious attention, even though cortical functions tend to be fractionated across numerous discrete brain regions (Treisman 1996; Noppeney 2021; Scholte and de Haan 2025). By assigning relevance values to information from different forebrain subsystems and then converting this information into a correlated pattern of metabolic support, the NME loop enables the possibility that local information indirectly influences the subsequent distribution of global neural information processes. Thus, independent of the way that metabolic differences are responsive to local neural activity, regional metabolism can also be modulated *non-locally*.

Because the multi-synaptic NME loop of connections converges on the relatively small midbrain and brainstem nuclei, this transition constitutes a connectional bottleneck (diagrammed in Figure 4). Each of these regulatory nuclei contains a comparatively small population of neurons compared to the structures that comprise their inputs and outputs. As a result, not only are patterns of neural information transformed into metabolic information at this juncture, but the complexity and diversity of this information are also maximally reduced to the up- and down-regulation of relatively simple metabolic parameters, such as oxygen delivery. This extensive compression of complex simultaneous brain processes offers a possible explanation of how the highly fractionated information processes of the cerebral cortex become integrated, which makes the vastly many relatively autonomous and unconscious brain functions become coordinated with one another to provide a globally integrated, and unified conscious experience.

Since these midbrain systems are also targets of blood-borne chemical signals, nonlocal modulation enables somatic conditions such as hunger, fatigue, arousal, and stress to also become integrated into the changes of metabolism that reconfigure global neural activity in response to changing conditions.

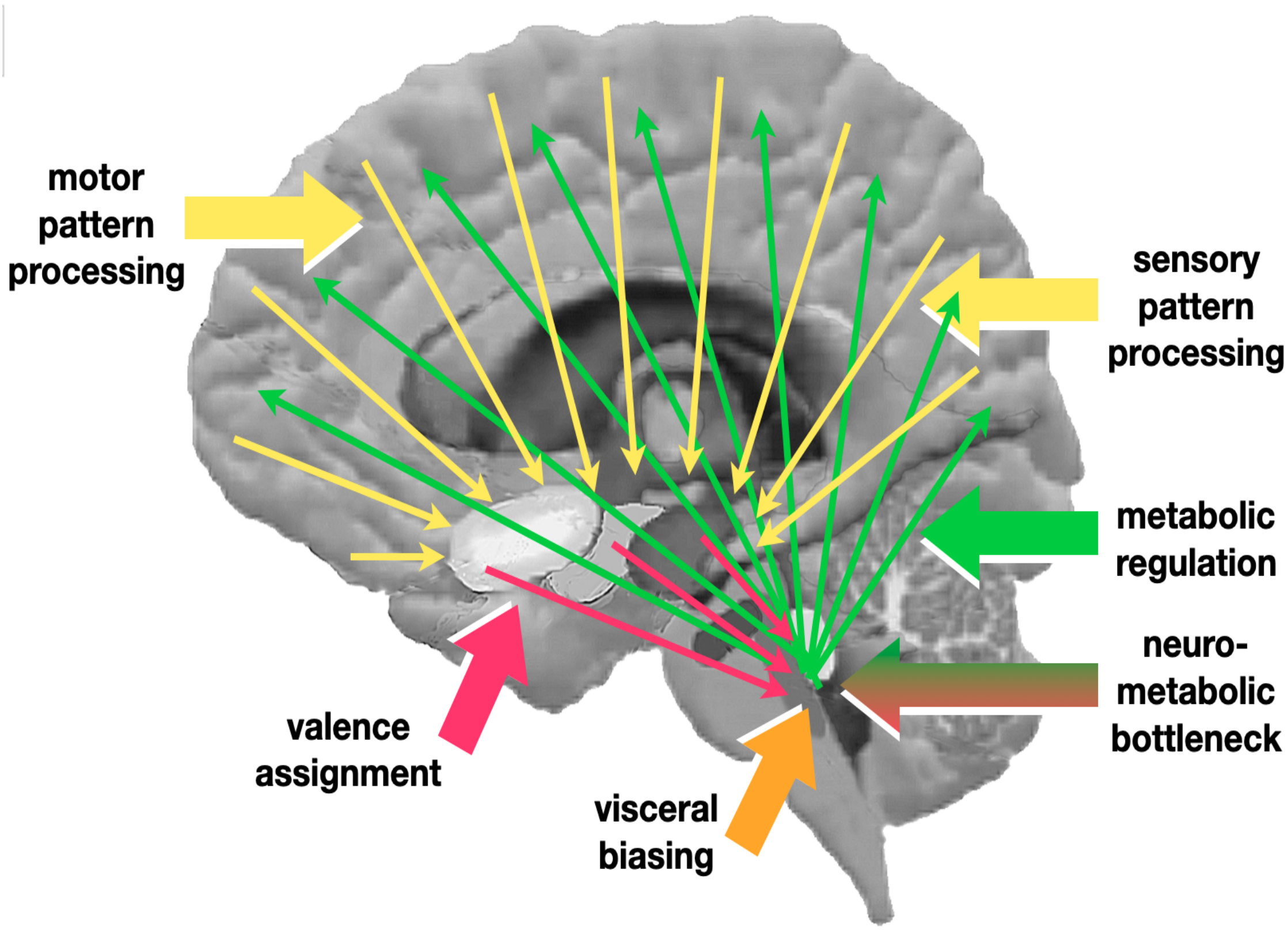

**Figure 4**. A depiction of the three-stage *neuro-metabolic bottleneck* that compresses high dimensional forebrain information into value-correlated re-distribution of forebrain physiological support, as follows. Neural pattern analysis in diverse forebrain regions is first relayed (yellow arrows) to basal forebrain nuclei. These basal forebrain nuclei compress this complex forebrain information into either attractive stimuli (likely beneficial = *approach*), repulsive stimuli (likely harmful = *avoid*), or indifferent stimuli. This information is then relayed (red arrows,) from these forebrain nuclei to midbrain-brainstem nuclei where this valence information is translated into correlated physiological-regulatory information. Neural outputs from these nuclei (green arrows) reach nearly all areas of the forebrain where they regulate the redistribution of energy and various physiological functions. (typical neurotransmitter molecules associated with each stage: yellow = glutamate, red = GABA, green = aromatic amines: norepinephrine, serotonin, dopamine).

This connectional bottleneck also includes temporal integration of experience as well. Over time, the preceding cycle of brain activity will shape the metabolic conditions within which the next cycle unfolds. Thus, each iteration of the neurometabolic loop operates on an already modified medium so that at any moment, the current brain state reflects the history of its past states and shapes its future states by changing the ground (the medium) rather than the figure (the message). Traced across many cycles and differentiated forebrain regions, this process becomes deeply recursive and the formal and material attributes of experience become ever more inextricably entangled.

## 6. The locus of experience

We propose that the neurometabolic entanglement created by the Möbius-like loop of neural connections can account for two central characteristics of the self: the emergence of a perspective and the experience of agency. We now examine each of these.

Biologists take it for granted that, starting with the simplest organisms, all living things are linked by an unbroken lineage of selfhood. Bacteria reproduce them*selves*; plants sustain them*selves* by growing toward sunlight; animals protect them*selves* by hiding from predators. We call this elementary form of selfhood that evolved billions of years before the first nervous systems the *somatic self*. In contrast, the conscious self is a highly specialized manifestation of this more general biological phenomenon—one shared by thousands of species, with or without brains. It is no accident that all known instances of phenomenal consciousness occur in organisms. In this respect, the conscious sense of self that emerges from the NME loop can be re-identified as a neural loop which stands in a double relationship to the larger somatic loop (the Möbius loop of the whole organism). On one side, the brain is a biological organ, and like any other organ, its metabolism depends on the body's metabolism. On the other, it re-embodies the somatic loop in the form of brain activity while regulating the very body of which it is a part. This relationship mirrors the Möbius-like entanglement already described within the brain itself, but at a higher level of organization: so that brain and body are similarly mutually entangled. The result is a nested hierarchy of loops within loops, such that the neurometabolic loop of the brain is embedded within, and continuously coupled to, the broader metabolic loop of the living body. This self-within-self nesting is depicted in Figure 5.

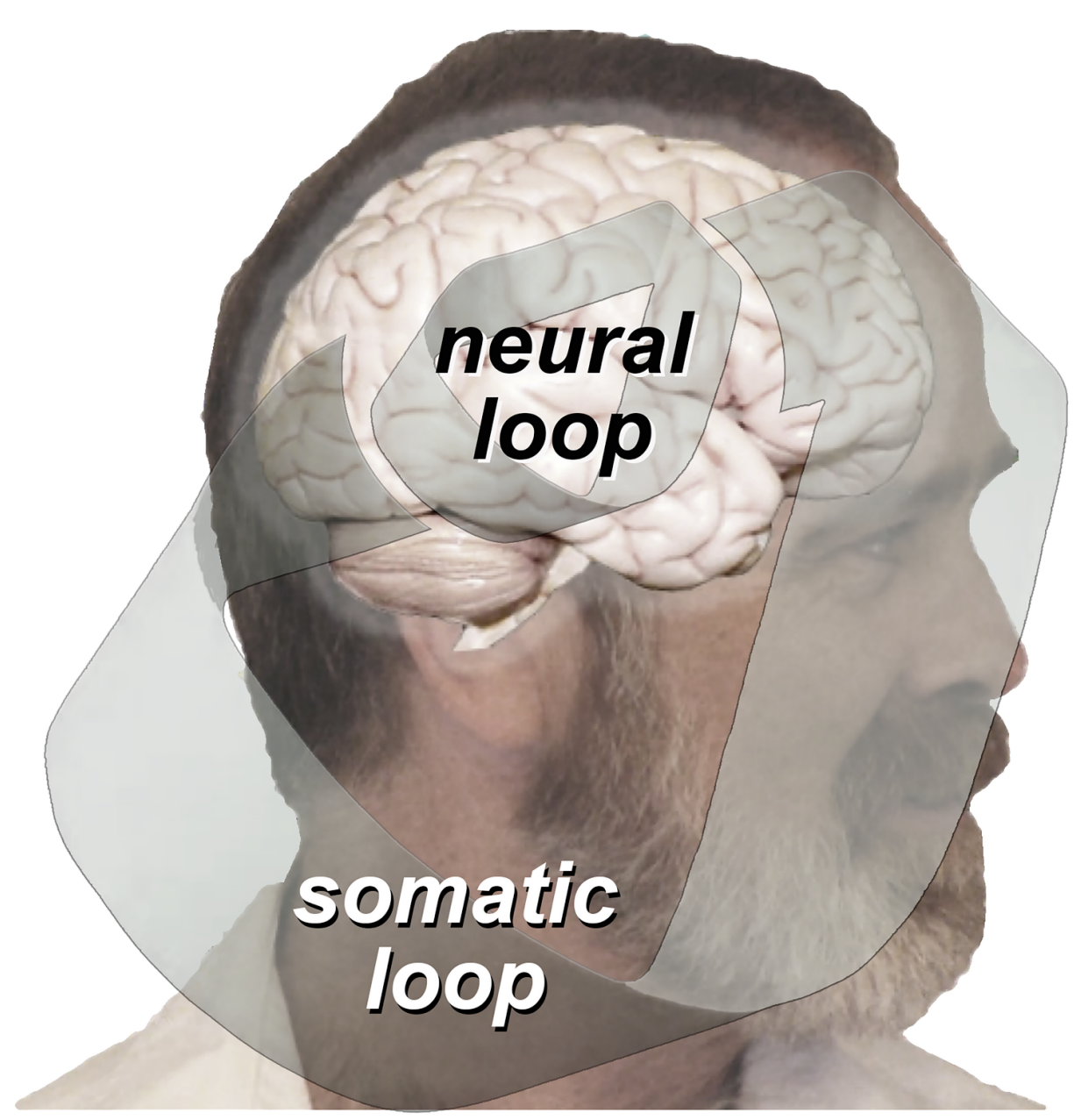

**Figure 5.** A depiction of the doubly nested entanglement of the neural Möbius loop within a higher order Möbius loop of the whole organism. This doubly recursive structure

displaces the neural loop from the somatic loop of which it is a part, so that its activity re-presents itself to itself. This displacement of the neural message from the neural-somatic medium with which it is entangled enables neural processes to re-present the ways that their physicality can be modified by interaction with the physical world.

By itself, somatic self-maintenance does not translate into phenomenal consciousness. For instance, the immune response and food digestion are expressions of the somatic self —both are self-directed, self-regenerating processes that distinguish self from non-self as beneficial or disruptive conditions. Yet the myriad of operations involved in digesting nutrients or neutralizing pathogens are not themselves experienced. The same applies to the vast majority of neural activity engaged in body regulation. This suggests that conscious selfhood—the neurological self—is a *displaced* elaboration of the somatic self. The question, then, is what distinguishes those forms of self-organization that become conscious from the many that do not. We believe that these are two: the emergence of a perspective and the directionality of agency.

*The emergence of perspective.* The most fundamental feature of conscious experience is not its qualitative character—its colors, felt efforts, or sensations of warmth—but the fact that it is always *someone's* experience. There is no such thing as experience from nowhere; "every experience is given from a particular point of view, one that cannot be reduced or exchanged for any other.

The NME loop can explain the perspectival character of experience. At each cycle, the loop generates a distinction between two things: the anticipatory system — the metabolic-neural state of the Möbius-like loop itself — and the neural activity pattern being anticipated. The current physiological-metabolic state of the system functions as a reference frame, that is, the condition against which incoming perturbations are measured. The current physiological-metabolic state also determines how the brain allocates its energetic resources in anticipation of change in the only sense that matters: as the stable-but-dynamic metabolic ground with respect to which all metabolic deviations are compared and from which all anticipatory neural modifications are initiated.

In this respect, the NME theory diverges from theories of consciousness (Dennett 1992; Metzinger 2003) that treat the self as an illusion. The self/nonself distinction is not a *model*; that is, it is not cognitively imposed on neural processing but an intrinsic feature of the brain's functional organization. What lies on one side of the metabolic boundary is experienced as *self*, while what impinges from outside it is experienced as *nonself.* This reflects the alternation between neural activity patterns influencing metabolism and metabolism influencing neural activity. Both are continuously modified, and their entanglement is cyclically regenerated to provide the unbroken continuity of the self.

At the same time, the NME loop never begins from zero. Each present state carries the trace of what preceded it as a *metabolic inheritance*: the particular distribution of excitability, sensitivity, and energetic readiness caused by the metabolic footprints of the prior cycle. Each present state also reaches forward into the next by organizing —rather than modelling— current metabolic allocation in anticipation of future dynamics.

Because each state is both the product of its predecessor and the condition of its successor, recursive self-modification creates what might be called an unbroken chain of metabolic *becoming*. The sense of continuity across time is the experiential form of this unbroken inheritance. And what feels like the elusiveness of the sense of self — the sense that the self cannot quite be apprehended, that it slips away when it is brought into focus— is precisely what we should expect from a reference frame that is never still, like a ground that is always shifting beneath the figures it sustains.

*The directionality of agency.* Not every conscious experience is accompanied by a sense of agency. While skilled action, deliberate thought, and imaginative engagement feel subjectively authored, a large swath of perceptual activity, surprise, and overwhelm feel externally received. Once a self exists as a metabolic reference frame, the following question arises: why do some experiences feel like something *I am doing*, while others feel like something happening *to me?* NME offers a specific and testable account of this experiential asymmetry.

The key lies in the *directionality* of metabolic reorganization within the NME loop. When the redistribution of metabolic work across the loop is driven primarily by its prior state —when the system reorganizes itself in anticipation of future perturbations— the reference frame internally guides the metabolic change. In such moments, the self is the source of the reorganization rather than its target, and the resulting experience carries the distinctive quality of agency. By contrast, when metabolic reorganization is driven by external perturbation —by stimuli the NME loop could neither anticipate nor have initiated— the change originates from outside the reference frame, and experience takes on a passive quality.

The general sense of agency is hence the experiential manifestation of the source of metabolic reorganization. This has a testable implication: conditions that increase the contribution of internally-generated predictive activity relative to externally driven perturbation such as skilled performance, focused attention, and deliberate imagination, should produce heightened feelings of agency, while conditions that maximize the gap between what was predicted and what actually arrives should produce the feeling of being overwhelmed, of events outrunning the self's capacity to meet them.

The emergence of a perspective and the directionality of agency are complementary expressions of an entangled self. In each case, the explanation does not invoke a separate mechanism or a higher-level perspective. It is the experiential form of the entanglement between information and its neural medium traced at a different level of description. Perspective is what the NME loop's reference frame *is*, from the inside. Agency is what it *feels like* when metabolic reorganization originates within that frame. The result is that neurometabolic entanglement renders neural activity intrinsically about its own physiological grounding.

## 7. Conclusion

A deeply entangled relationship between neural information processes and metabolic modification of the neuronal medium helps to account for the range of ways that conscious experience can be extrinsically manipulated. For example, dissociations in neurovascular changes and neuronal activity appear to underlie the total loss of consciousness under general anesthesia (Wu et al. 2019, Rungta et al. 2021, Wang et al. 2024a). Psychedelic drugs likewise dissociate neurovascular coupling (Carhart-Harris et al. 2012, Lewis et al. 2017, Larsen et al. 2025, Padawer-Curry et al. 2025), indicating that it is not only the level of consciousness but the content of phenomenal experience that is affected. Spontaneous transitions, like the transition from sleep to wakefulness, are likewise accompanied by reductions in the magnitude of, and lags in the timing of, vascular coupling to neural signals (Winder et al. 2017, Turner et al. 2020, Natsubori et al. 2024, Jacob et al. 2025, Landemard et al. 2026).

As also exemplified by other experiential phenomena, such as metabolic correlations with the experience of mental effort or arousal intensity, these predictions demonstrate that the theory of neuro-metabolic entanglement is empirically testable and its experiential, functional, and anatomical claims are falsifiable. This also provides grounds for further empirical investigation. For example, it is consistent with the trend toward refocusing psychological health beyond merely analyzing thought and behavior patterns and toward a better understanding of the metabolic conditions that underlie disturbances of mental function. In this respect, it should not be surprising that nearly all current neuropsychiatric drugs and consciousness-altering compounds influence midbrain-brainstem regulatory neurotransmitter systems. The possibility of reframing our understanding of brain activity to empirically investigate these and other unexplored forms of entangled relationships makes NME an empirically generative alternative to most theories of consciousness—many of which are deemed unfalsifiable.

In conclusion, once we abandon the assumption that the physical and experiential properties are independently characterizable phenomena, the question of why a physical process can produce an experiential quality loses its force. And by demonstrating how these processes are neurologically entangled, aspects of this hypothesis are subject to empirical test. In summary, the theory of neuro-metabolic entanglement does not claim to have solved the "hard problem" of consciousness, nor does it offer a bridge across any explanatory gap. Instead, it demonstrates that such a gap does not exist.

**References**

Bisenius, S., Trapp, S., Neumann, J., & Schroeter, M. L. (2015). Identifying neural correlates of visual consciousness with ALE meta-analyses. *NeuroImage*, *122*, 177–187.

Buxton, R. B. (2012). Dynamic models of BOLD contrast. *NeuroImage*, *62*(2), 953–961.

Carhart-Harris, R. L., Erritzoe, D., Williams, T., Stone, J. M., Reed, L. J., Colasanti, A., Tyacke, R. J., Leech, R., Malizia, A. L., Murphy, K., Hobden, P., Evans, J., Feilding, A., Wise, R. G., & Nutt, D. J. (2012). Neural correlates of the psychedelic state as determined by fMRI studies with psilocybin. *Proceedings of the National Academy of Sciences of the United States of America*, *109*(6), 2138–2143.

Dennett, C. (1992). *Consciousness Explained*. Little, Brown & Company.

Drew, P. J. (2019). Vascular and neural basis of the BOLD signal. *Current Opinion in Neurobiology*, *58*, 61–69.

Francken, J. C., Beerendonk, L., Molenaar, D., Fahrenfort, J. J., Kiverstein, J. D., Seth, A. K., & van Gaal, S. (2022). An academic survey on theoretical foundations, common assumptions and the current state of consciousness science. *Neuroscience of Consciousness*, *2022*(1), niac011.

Friston, K. (2010). The free-energy principle: a unified brain theory? *Nature Reviews. Neuroscience*, *11*(2), 127–138.

Giorgi, F. S., Galgani, A., Puglisi-Allegra, S., Limanaqi, F., Busceti, C. L., & Fornai, F. (2020). Locus Coeruleus and neurovascular unit: From its role in physiology to its potential role in Alzheimer's disease pathogenesis. *Journal of Neuroscience Research*, *98*(12), 2406–2434.

Hesselmann, G., & Malach, R. (2011). The link between fMRI-BOLD activation and perceptual awareness is "stream-invariant" in the human visual system. *Cerebral Cortex (New York, N.Y.: 1991)*, *21*(12), 2829–2837.

Hillman, E. M. C. (2014). Coupling mechanism and significance of the BOLD signal: a status report. *Annual Review of Neuroscience*, *37*, 161–181.

Howarth, C., Mishra, A., & Hall, C. N. (2021). More than just summed neuronal activity: how multiple cell types shape the BOLD response. *Philosophical Transactions of the Royal Society of London. Series B, Biological Sciences*, *376*(1815), 20190630.

Jacob M, Ford J, Deacon T. 2023. Cognition is entangled with metabolism: relevance for resting-state EEG-fMRI. *Frontiers in human neuroscience 17*: 976036.

Jacob MS, Ford JM, Roach BJ, Calhoun VD, Mathalon DH. 2019. Aberrant activity in conceptual networks underlies N400 deficits and unusual thoughts in schizophrenia. *NeuroImage: Clinical* 101960.

Jacob MS, Roach BJ, Mathalon DH, Ford JM. 2025. Noncanonical neural-hemodynamic coupling by default and in schizophrenia. *Biological Psychiatry: Cognitive Neuroscience and Neuroimaging.*

Korte, N., James, G., You, H., Hirunpattarasilp, C., Christie, I., Sethi, H., & Attwell, D. (2023). Noradrenaline released from locus coeruleus axons contracts cerebral capillary pericytes via α2 adrenergic receptors. *Journal of Cerebral Blood Flow and Metabolism: Official Journal of the International Society of Cerebral Blood Flow and Metabolism*, *43*(7), 1142–1152.

Landemard, A., Krumin, M., Harris, K. D., & Carandini, M. (2026). Brainwide blood volume reflects opposing neural populations. *Nature*. https://doi.org/10.1038/s41586-026-10350-9

Larsen, K., Lindberg, U., Ozenne, B., McCulloch, D. E., Armand, S., Madsen, M. K., Johansen, A., Stenbæk, D. S., Knudsen, G. M., & Fisher, P. M. (2025). Acute psilocybin and ketanserin effects on cerebral blood flow: 5-HT2AR neuromodulation in healthy humans. *Journal of*

*Cerebral Blood Flow and Metabolism: Official Journal of the International Society of Cerebral Blood Flow and Metabolism*, *45*(7), 1385–1401.

Lecrux, C., Bourourou, M., & Hamel, E. (2019). How reliable is cerebral blood flow to map changes in neuronal activity? *Autonomic Neuroscience: Basic & Clinical*, *217*, 71–79.

Lewis, C. R., Preller, K. H., Kraehenmann, R., Michels, L., Staempfli, P., & Vollenweider, F. X. (2017). Two dose investigation of the 5-HT-agonist psilocybin on relative and global cerebral blood flow. *NeuroImage*, *159*, 70–78.

Logothetis, N. K., & Wandell, B. A. (2004). Interpreting the BOLD signal. *Annual Review of Physiology*, *66*, 735–769.

Magistretti, P. J., & Allaman, I. (2015). A cellular perspective on brain energy metabolism and functional imaging. *Neuron*, *86*(4), 883–901.

Metzinger, T. (2003). *Being no one: The self-model theory of subjectivity*. https://direct.mit.edu/books/monograph/1991/Being-No-OneThe-Self-Model-Theory-of-Subjectivity

Meyer-Baese, L., Jaeger, D., & Keilholz, S. (2025). Neurovascular coupling: a review of spontaneous neocortical dynamics linking neuronal activity to hemodynamics and what we have learned from the rodent brain. *Journal of Neurophysiology*, *133*(2), 644–660.

Morgane, P. J., Galler, J. R., & Mokler, D. J. (2005). A review of systems and networks of the limbic forebrain/limbic midbrain. *Progress in Neurobiology*, *75*(2), 143–160.

Mu, X., Yan, J., Zheng, Y., Jiang, X., Weng, N., Yao, L., Sha, W., & Chen, Y. (2026). Brainstem raphe nuclei and their serotonergic neurons: An integrative review of distribution, function and projection. *Behavioural Brain Research*, *515*(116415), 116415.

Müller, M. J., Wang, Z., Heymsfield, S. B., Schautz, B., & Bosy-Westphal, A. (2013). Advances in the understanding of specific metabolic rates of major organs and tissues in humans. *Current Opinion in Clinical Nutrition and Metabolic Care*, *16*(5), 501–508.

Natsubori, A., Kwon, S., Honda, Y., Kojima, T., Karashima, A., Masamoto, K., & Honda, M. (2024). Serotonergic regulation of cortical neurovascular coupling and hemodynamics upon awakening from sleep in mice. *Journal of Cerebral Blood Flow and Metabolism: Official Journal of the International Society of Cerebral Blood Flow and Metabolism*, *44*(9), 1591–1607.

Noppeney, U. (2021). Perceptual inference, learning, and attention in a multisensory world. *Annual Review of Neuroscience*, *44*(1), 449–473.

Özçete, Ö. D., Banerjee, A., & Kaeser, P. S. (2024). Mechanisms of neuromodulatory volume transmission. *Molecular Psychiatry*, *29*(11), 3680–3693.

Padawer-Curry, J. A., Krentzman, O. J., Kuo, C.-C., Wang, X., Bice, A. R., Nicol, G. E., Snyder, A. Z., Siegel, J. S., McCall, J. G., & Bauer, A. Q. (2025). Psychedelic 5-HT2A receptor agonism alters neurovascular coupling and differentially affects neuronal and hemodynamic measures of brain function. *Nature Neuroscience*, *28*(11), 2330–2343.

Pessoa, L. (2024). The spiraling cognitive-emotional brain: Combinatorial, reciprocal, and reentrant macro-organization. *Journal of Cognitive Neuroscience*, *36*(12), 2697–2711.

Poldrack, R. A., Sabb, F. W., Foerde, K., Tom, S. M., Asarnow, R. F., Bookheimer, S. Y., & Knowlton, B. J. (2005). The neural correlates of motor skill automaticity. *The Journal of Neuroscience: The Official Journal of the Society for Neuroscience*, *25*(22), 5356–5364.

Raichle, M. E. (2015). The restless brain: how intrinsic activity organizes brain function. *Philosophical Transactions of the Royal Society of London. Series B, Biological Sciences*, *370*(1668). https://doi.org/10.1098/rstb.2014.0172

Raichle, M. E., & Gusnard, D. A. (2002). Appraising the brain's energy budget. *Proceedings of the National Academy of Sciences of the United States of America*, *99*(16), 10237–10239.

Rungta, R. L., Zuend, M., Aydin, A.-K., Martineau, É., Boido, D., Weber, B., & Charpak, S. (2021). Diversity of neurovascular coupling dynamics along vascular arbors in layer II/III somatosensory cortex. *Communications Biology*, *4*(1), 855.

Scholte, H. S., & de Haan, E. H. F. (2025). Beyond binding: from modular to natural vision. *Trends in Cognitive Sciences*, *29*(6), 505–515.

Seth, A. K., & Bayne, T. (2022). Theories of consciousness. *Nature Reviews. Neuroscience*, *23*(7), 439–452.

Sun, L., Peng, X.-L., Zi, H.-X., Zhang, Z.-K., Gong, Y.-C., Li, J., Zhu, Y.-Q., Yang, Y.-Q., Cao, H.-W., Hu, D., Wang, K., & Du, J.-L. (2026). Globally patterned locus coeruleus-norepinephrine neuron-pericyte coupling orchestrates brain-wide vascular dynamics. *Neuron*, *114*(2), 287-306.e9.

Theriault, J. E., Katsumi, Y., Reimann, H. M., Zhang, J., Deming, P., Dickerson, B. C., Quigley, K. S., & Barrett, L. F. (2025). It's not the thought that counts: Allostasis at the core of brain function. *Neuron*, *113*(24), 4107–4133.

Tononi, G., Boly, M., Massimini, M., & Koch, C. (2016). Integrated information theory: from consciousness to its physical substrate. *Nature Reviews. Neuroscience*, *17*(7), 450–461.

Treisman, A. (1996). The binding problem. *Current Opinion in Neurobiology*, *6*(2), 171–178.

Turner, K. L., Gheres, K. W., Proctor, E. A., & Drew, P. J. (2020). Neurovascular coupling and bilateral connectivity during NREM and REM sleep. *eLife*, *9*. https://doi.org/10.7554/eLife.62071

Uddin, L. Q. (2020). Bring the Noise: Reconceptualizing Spontaneous Neural Activity. *Trends in Cognitive Sciences*. https://doi.org/10.1016/j.tics.2020.06.003

Wang, X., Padawer-Curry, J. A., Bice, A. R., Kim, B., Rosenthal, Z. P., Lee, J.-M., Goyal, M. S., Macauley, S. L., & Bauer, A. Q. (2024). Spatiotemporal relationships between neuronal, metabolic, and hemodynamic signals in the awake and anesthetized mouse brain. *Cell Reports*, *43*(9), 114723.

Winder, A. T., Echagarruga, C., Zhang, Q., & Drew, P. J. (2017). Weak correlations between hemodynamic signals and ongoing neural activity during the resting state. *Nature Neuroscience*, *20*(12), 1761–1769.

Wohlschläger, A. M., Glim, S., Shao, J., Draheim, J., Köhler, L., Lourenço, S., Riedl, V., & Sorg, C. (2016). Ongoing Slow Fluctuations in V1 Impact on Visual Perception. *Frontiers in Human Neuroscience*, *10*, 411.

Wu, G.-R., Di Perri, C., Charland-Verville, V., Martial, C., Carrière, M., Vanhaudenhuyse, A., Laureys, S., & Marinazzo, D. (2019). Modulation of the spontaneous hemodynamic response function across levels of consciousness. *NeuroImage*, *200*, 450–459.

**Statement of author contributions:**

All three authors contributed to the research and writing of this article.

**Conflicts of interest statement:**

We declare no conflicts of interest

**Funding statement:**

Work on this article by Miguel Garcia-Valdecasas was funded by Ministerio de Ciencia, Innovación y Universidades (project PID2022-140659NB-I00).

Work on this article by Michael Jacob was funded by grants from the Department of Veterans Affairs (IK2CX002457)

**Acknowledgements:** Thanks to members of the *Terry and the Pirates* discussion group for reading and commenting on earlier drafts of this manuscript.